\documentclass[aps,prd]{revtex4}

\usepackage{amssymb}
\usepackage{epsfig}
\usepackage{graphicx,psfrag}
\usepackage{bm}
\newcommand{\be}{\begin{equation}}
\newcommand{\ee}{\end{equation}}
\newcommand{\bea}{\begin{eqnarray}}
\newcommand{\eea}{\end{eqnarray}}
\newcommand{\hf}{\frac12}

\def\journal#1#2#3#4{{\em#1} {\bf#2}, #4 (#3)}

\def\eq#1{(\ref{#1})}

\def\la{\langle}
\def\ra{\rangle}
\def\ord#1{{\cal O}(#1)}
\def\dk{\Delta k}

\begin{document}
\title{Gluon confinement and quantum censorship\footnote{Talk prepared for the
conference ``Gribov-80``, May 26-28, 2010, ICTP Trieste, Italy}}

\author{Janos Polonyi}

\affiliation{Strasbourg University, High Energy Theory Group, CNRS-IPHC,\\
23 rue du Loess, BP28 67037 Strasbourg Cedex 2 France\\
E-mail: polonyi@ires.in2p3.fr}

\begin{abstract}
The dynamical Maxwell-cut, a degeneracy is shown to be a precursor 
of condensate in the $\phi^4$ and the sine-Gordon models. The difference
of the way the Maxwell-cut is obtained is pointed out and quantum censorship, the
generation of semiclassically looking phenomenon by loop-corrections is conjectured
in the sine-Gordon model. It is argued that quantum censorship and gluon confinement
exclude each other.
\end{abstract}


\maketitle

\section{Introduction}
The peculiarity of a condensate arises from its macroscopic occupation.
The order of magnitude of the typical quantum fluctuations is $\ord{\hbar^{1/2}}$
in a trivial, perturbative vacuum. When a condensate is present then the field 
is supposed to display an $\hbar$-independent, $\ord{\hbar^0}$ expectation value. 
The expectation value of the square of the field in a state with $n$ particles is $\hbar\ord{n}$
thus an occupation number $\ord{\hbar^{-1}}$ is needed to make up the condensate.

The macroscopically high occupation number is possible in the absence of 
strong repulsive forces acting between the particles only. In other words, the presence of
a condensate suggests a high degree of degeneracy in the vacuum, a necessary condition
of the semiclassical approximation. A dynamical generalization of the Maxwell-cut 
was found by inspecting the semiclassical contributions to the functional
renormalization group equations in theories with condensate\cite{jean}. It turned out that the loop contributions
may change this picture and quantum censorship was proposed as a mechanism which 
reduces a semiclassical degeneracy to an approximate degeneracy driven by quantum 
fluctuations\cite{vincentphi,vincentsg}. 

The high degree of degeneracy indicating the emerging condensate influences the 
asymptotic sector of a theory in a fundamental manner. When created semiclassically
then the degeneracy consists of localized modes. In the quantum censorship
scenario the loop contributions are handled in the extended plane wave basis and 
their way of realizing the degeneracy is based on extended modes. The Yang-Mills 
vacuum contains a condensate, as well, and the structure of its soft is obviously 
important for confinement. It is argued below that gluon confinement excludes 
the building up of the quantum censorship.

\section{Tree-level condensates}
We consider condensate in a weakly coupled Euclidean field theory for a real scalar 
field $\phi(x)$ where the sharp UV cutoff in momentum space will be lowered by the
recursive equation
\be\label{block}
e^{-\frac1\hbar S_{k-\dk}[\phi]}=\int D[\phi']e^{-\frac1\hbar S_{k-\dk}[\phi+\phi']}.
\ee
Here $\phi(x)$ and $\phi'(x)$ denote fields with support in $0<p<k-\dk$ and 
$k-dk<p<k$, respectively. We follow for simplicity the local potential approximation where
the form
\be\label{ansatz}
S_k[\phi]=\int dx\left[\hf(\partial\phi)^2+U_k(\phi)\right]
\ee
is assumed. Most of the examples mentioned below correspond to the initial condition
$U_{\Lambda}(\phi)=m^2_B\phi^2/2+g_B\phi^4/4!$ in four space-time dimensions. The functional integral will be 
evaluated in the loop-expansion and the Wegner-Houghton equation\cite{wh} 
\be\label{wh}
\partial_kU_k(\phi)=-\frac{\hbar k^3}{16\pi^2}\ln[k^2+U_k''(\phi)],
\ee
follows by assuming trivial saddle points. This is an exact equation
because the small parameter of the loop-expansion is actually $\hbar\dk/k$ because each
loop integral is restricted to a shell of thickness $\dk$ in momentum space. But what
happens when nontrivial saddle points are encountered? 

\begin{figure}
\psfig{file=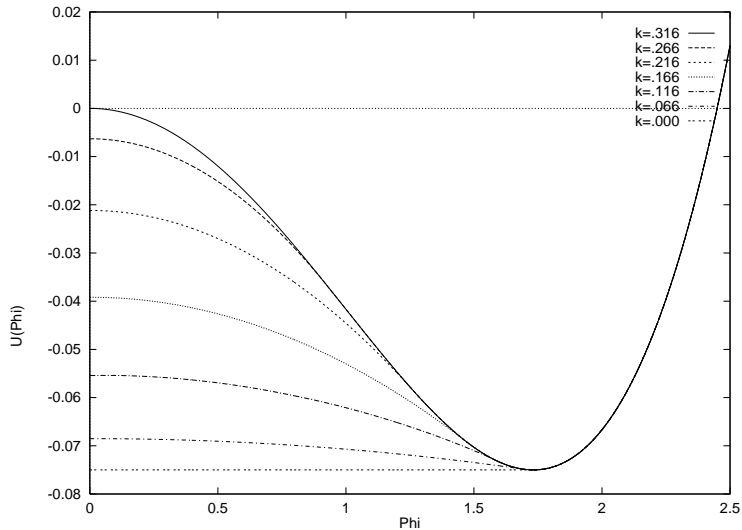,width=7cm}
\caption{Evolution of the potential \cite{jean} in the unstable region $|\Phi|<\Phi_k$
for $m_B^2=-0.1$, $g_B=0.2$ in units of $k_{init}=\Lambda=1$.}
\label{treemaxw}
\end{figure}

The simplest strategy to find an answer is to ignore the loop corrections altogether 
and consider the tree-level evolution only \cite{jean}. It turns out that the
nontrivial saddle point of the blocking relation \eq{block} is the hallmark of
spontaneous symmetry breaking, a condensate in the vacuum. 
There is no evolution on the tree-level as long as the action reaches its minimum at $\phi'(x)=0$
within the shell $k-dk<p<k$. But if the potential has negative curvature at the initial 
condition at $k=\Lambda$ say at $\phi=0$ then the gradual decrease of the gliding cutoff 
$k$ leads to nontrivial saddle points at some finite scale $k=k_{cr}$. The evolution
from this scale was followed numerically by setting up an iterative process where
the cutoff was decreased by a small but finite amount, $k\to k-\dk$ and for each value of the
homogeneous field configuration $\Phi$ a saddle point $\phi'_{cl}$ was sought within the 
family of plane waves with momentum $k$ and the new potential 
$VU_{k-\dk}(\Phi)=S_k[\Phi+\phi'_{cl}]$ was calculated. The potential evolved in an interval
around zero, $|\phi|<\Phi_k$ where $\Phi_k$ increases from zero as $k$ decreases from
$k_{cr}$ and the form 
\be\label{instpot}
V^{inst}_k(\phi)=-\frac{k^2}2\phi^2
\ee
was found as shown in Fig. \ref{treemaxw}. 

Such a potential makes the action \eq{ansatz} degenerate at the scale $p=k$ by canceling the kinetic term in. What
happened numerically was that the action lost convexity at $k=k_{cr}$ and in the next step,
$k\to k_{cr}-\dk$ a shallow, $\ord{\dk/k}$ minimum was found in the action as the function of
the plane wave amplitude. This saddle point made the action $S_{k_{cr}-\dk}[\phi]$
degenerate for modes with $p=k-\dk$ with a slightly larger amplitude. This scenario
repeated itself during the iteration, namely the saddle point, found in a shallow
minimum made the new action degenerate again. The best description of the dynamics of
a mode with a given momentum $p$ is provided by $S_k[\phi]$ with $k=p$, hence the
degeneracy of the blocked actions at their cutoff can be interpreted as the indication
within the renormalization group scheme that the dynamics is degenerate for all modes with
small enough amplitude and momentum $p<k_{cr}$. This is a dynamical Maxwell cut, a precursor of the condensate, 
and leads to the traditional flattening of the effective potential at $k=0$,
$U_{eff}(\Phi)=V_{k=0}(\Phi)$ between the vacuum expectation values 
$\la0|\phi(x)|0\ra=\pm\Phi_{vac}$. It is natural to interpret the plane wave saddle points 
as domain walls. There is a saddle point for each scale $p<k_{cr}$ and the functional 
integration over their soft zero modes, arising from the their breaking of the space-time
symmetries makes the system strongly coupled as expected by analogies with the
mixed phase of first order phase transitions.

Though these results are simple and reasonable, they raise some doubt and concern. 
The first question concerns the role of loop-correction. Can they alter the conclusion in a 
qualitative manner? The procedure is based on the sharp cutoff which prevents us from including
higher orders of the gradient expansion in the action \eq{ansatz}. Will the general picture
remain valid when these terms are taken into account in some manner?
The idea of the renormalization group is the construction of the renormalized, full
action by integrating a differential equation. The emergence of a saddle point 
usually signals the appearance of non-analytic terms. Will the evolution equation
remain integrable in $k$ at $k_{cr}$? 

As of the trouble spots, the first is related to the shallowness of the minimum of 
the action around the saddle point because it removes the factor $\dk/k$ from the small parameter 
of the loop expansion and we fall back on the $\hbar$-expansion. To make things worse,
we even loose the $\hbar$-expansion in the degenerate limit, $\dk\to0$, having 
a flat integrand. The evolution equation \eq{wh} requires the availability of the loop
expansion even if the higher orders are resummed in the differential equation limit 
$\dk\to0$. In general, we have neither analytical nor numerical method at hand to 
tackle constant integrands, the ultimate strong coupling limit.

Another troubling aspect of the soft zero mode dynamics of the domain walls  is that
it renders the physics of the mixed phase, $|\la0|\phi(x)|0\ra|<\Phi_{vac}$ non-perturbative 
and opens the possibility of having new free, relevant parameter in the scalar model,
such as the value of the condensate\cite{grg}.
The true vacuum lies at the common point of the stable region $\la0|\phi(x)|0\ra>\Phi_{vac}$
and the mixed phase therefore ``half'' of the quantum fluctuations which tend to decrease
$\phi(x)$ might well be non-perturbative.

\section{Loop-corrections}
\begin{figure}
\psfig{file=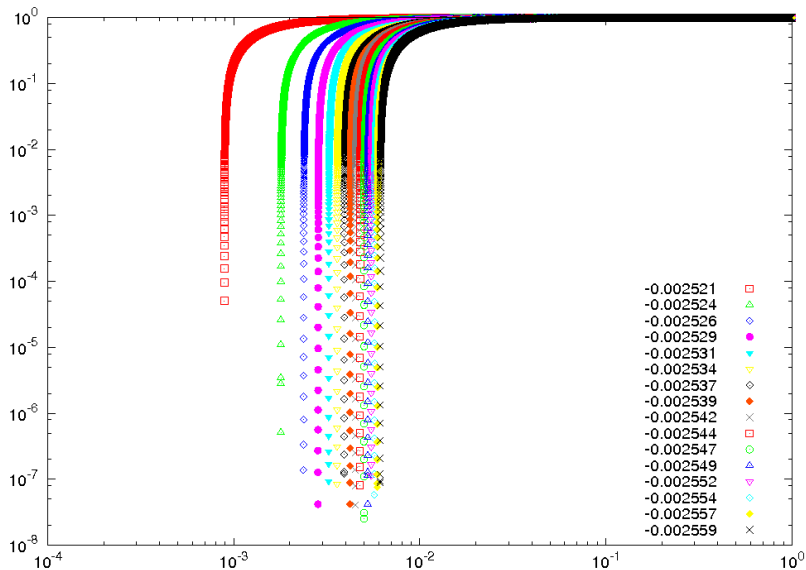,angle=270,width=11cm}
\caption{Dependence of the dimensionless degeneracy for vanishing field amplitude,
$P_k(0)$ of the action on the scale $k$ for $g_B=0.8$, the bare mass square values are 
shown in the figure \cite{vincentphi}.}
\label{deg}
\end{figure}

\begin{figure}
\psfig{file=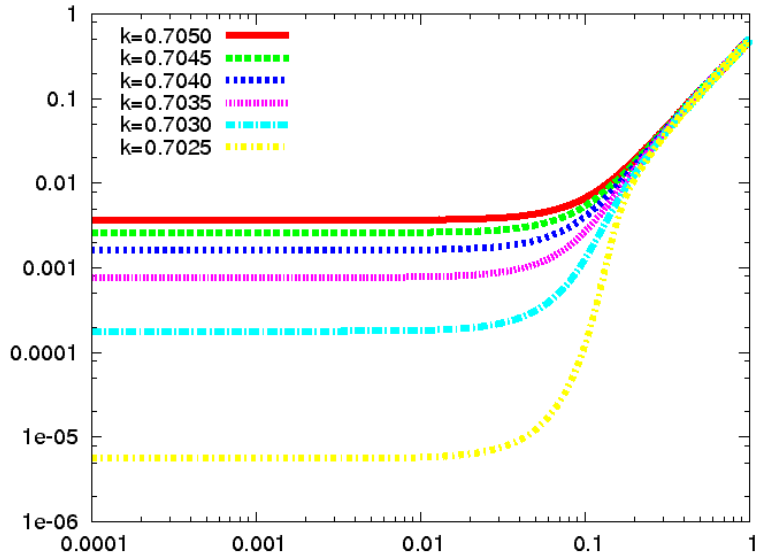,angle=270,width=11cm}
\caption{Evolution \cite{vincentphi} of $P_k(\Phi)$ as the function of the field 
$\Phi$ for $m_B^2=-0.5$ and $g_B=0.5$.}
\label{loopmaxw}
\end{figure}

The first of the previous questions is taken up in this section only, the eventual modification of the 
dynamical Maxwell-cut by including the loop corrections within the ansatz \eq{ansatz}. 
The potential is non-analytical according to the tree-level analysis therefore the
numerical integration of Eq. \eq{wh} must be done without assuming a polynomial
representation. This is possible by using the spline representation for the
potential \cite{vincentphi}. It is interesting to follow the evolution
of the dimensionless curvature of the action \eq{ansatz} for a mode $p=k$,
$P_k(\Phi)=1+\partial_\phi^2U_k(\Phi)/k^2$, depicted in Fig. \ref{deg}. It shows
a sudden drop at a finite scale followed by the stop of the program. The numerical 
algorithm makes dynamical adjustment of the precision needed during the integration
and stops because the system of linear equations for the spline coefficients becomes
nearly singular and requires extreme precision.
The value of $k$ where this happens depends slightly on the accuracy but seemed
to be not moving significantly up to our limit, several thousand splines. 
The potential approaches the form \eq{instpot} at the last steps as can clearly be seen from
Fig. \ref{loopmaxw}. The lesson of such a failed attempt to reach the infrared end point 
suggests a full or nearly complete dynamical Maxwell-cut.

The possible problems of the loop-expansion for a nearly degenerate action is avoided
when the evolution of the effective action is followed \cite{nicoll,wetterich,morris},
an algorithm with $\dk/k$ as the only small parameter. Similar, sudden drop of the
curvature of the action was observed at finite scale in this scheme when smooth cutoff
was used\cite{vincentphi}, leaving room for the eventual inclusion of higher orders of the
gradient expansion in the ansatz \eq{ansatz}. Naturally the difficulties of the degenerate
bare action remain in this scheme in the disguise of the problem of justifying any
usable ansatz for the effective action.

Can we decide whether the action is exactly or only nearly degenerate? The analytical efforts
presented so far\cite{ringwald,tetradis,wett,lipa} leave room for reaching true singularities
during integrating the evolution equation in the form of degenerate action involving 
non-analytical terms\cite{vincentsg}. Numerical methods, based on  finite amount of computer power
can give no satisfactory answer neither. I believe that the following options are left open:

\begin{itemize}
\item The evolution equation driven by loop-contributions leads to a truly degenerate action
as in the tree-level case. We have no analytical or numerical methods to tackle such
models.

\item The loop-contributions manage to keep the action regular and generate an approximative
dynamical Maxwell-cut. Such a shielding of the semiclassical singularity by 
quantum fluctuations which make up a similar effect is called Quantum Censorship.

\end{itemize}

\begin{figure}
\psfig{file=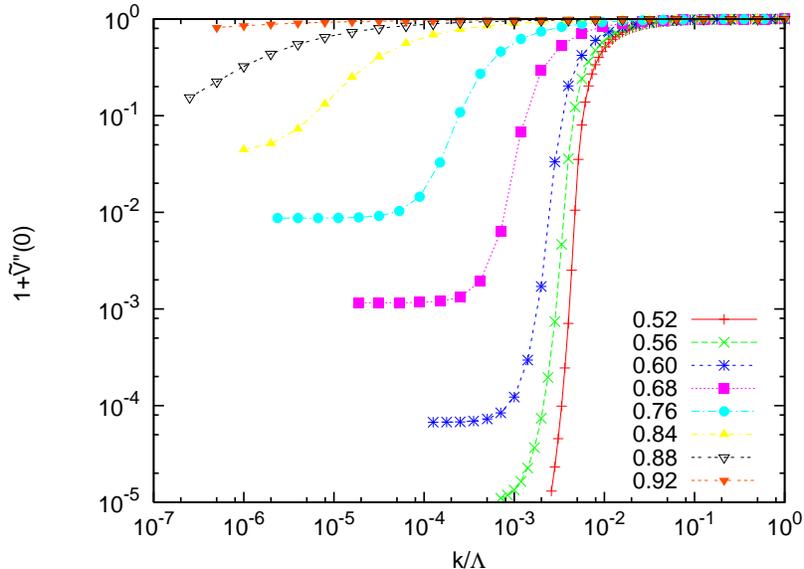,angle=270,width=11cm}
\caption{Evolution \cite{vincentsg} of $P_k(0)$ in the sine-Gordon model for different
values of $\beta_r$, indicated in the figure.for $m_B^2=-0.5$ and $g_B=0.5$.}
\label{sgdeg}
\end{figure}

Quantum Censorship seems to be realized in the two-dimensional sine-Gordon
model\cite{vincentsg} defined by the bare dimensionless potential 
$\tilde U_{k=\Lambda}(\phi)=k^{-2}U_B(\phi)=\tilde u_B\cos(\sqrt{8\pi}\beta_r\phi)$.
The effective potential must be constant, being the only convex periodic function, but
this naturally does prevent the model to display highly nontrivial dynamics in the 
infrared. The tree-level evolution gives saddle points and produces degenerate 
action\cite{nps} for $k<k_{cr}\ne0$ in the phase $\beta_r<1$. When the loop-contributions
are added then we observe a sudden drop of the curvature of the action, characteristic
of the dynamical Maxwell-cut followed by a surprising stabilization of the curvature 
at very small values\cite{vincentsg}.

\section{Yang-Mills theories}
We now leave the territory of well established results and make an attempt to interpret 
the difference of Figs. \ref{loopmaxw} and \ref{sgdeg} and to conjecture about 
its relevance for Yang-Mills theories.

It seems reasonable to assume that the increase of the accuracy of the numerical 
algorithm would find a plateau in Fig. \ref{sgdeg} for $\beta_r<0.6$, too.
When this state of affairs is compared with Fig. \ref{loopmaxw} which follows the evolution 
of models from nearly at the critical point to deep into the symmetry broken phase with the 
same numerical accuracy then one has the impression that Quantum Censorship is not observed 
in the non-periodic $\phi^4$ model. 

Accepting this interpretation one wonders about
the possible source of this difference between the two models. I believe that it is
to be found in the non-propagating nature of the excitations in the mixed phase
of the $\phi^4$ model. The domain walls formed in a vacuum with 
$|\la0|\phi(x)|0\ra|<\Phi_{vac}$ can be deformed with small energy
investment. As mentioned before, such a deformation represent the Goldstone modes 
of the broken space-time symmetries. These modes restore the homogeneity of
the vacuum like in a liquid but leave an important imprint on the dynamics by generating dissipation.
In the language of solid state physics the sound wave of the mixed phase is 
damped by the reflection and the traverse of a domain walls and the velocity of 
sound is reduced to zero. In particle physics this situation
is interpreted as having no asymptotic particle states. The domain
walls in the small $\beta_r$ phase of two dimensional sine-Gordon model where the
periodic symmetry of the theory is dynamically broken\cite{nps} are
kinks, stable propagating particles. As a result, elementary plane wave excitations have more chance to propagate
and the long distance structure of the theory can be reconstructed by means of
plane waves. Once this is possible the gradual turning on the plane waves during
the evolution may lead us to the correct vacuum.

Let us no turn to Yang-Mills theories. The vacuum of the scalar $\phi^4$ model 
lies at the boundary of the non-perturbative mixed phase and can safely be 
approached from the stable region. This is not the case in Yang-Mills theories where 
the vacuum is within the mixed phase. In fact, the strong chromo-magnetic attraction 
among gluons is supposed to generated a liquid vacuum filled with 
condensate\cite{savvidy} which is inhomogeneous\cite{nielsen,ken}, an analogy of the
mixed phase of the $\phi^4$ model. The conjecture that the integration of the
evolution equation of Yang-Mills theories in the plane wave basis would run into a
``naked singularity`` seems reasonable when gluons are confinement.
In fact, otherwise there should be colored asymptotic states. Though confinement excludes
strictly particle-like asymptotic gluon states only, non-particle like asymptotic 
states are not expected to exists neither according to the general view. In other words, color 
confinement and Quantum Censorship are exclusive properties.
The verification of this conjecture is a challenging question because I think
that we lack some technical elements related to gauge symmetry, Wick rotation and
the ansatz for the effective action.

There have been impressive advances achieved in applying the functional renormalization
group for Yang-Mills theories\cite{jan} by relying on gauge fixing and modified 
Slavnov-Taylor identities. A different method which avoids gauge fixing altogether 
or is explicitly independent of it would be useful to assure that the singularities 
arising from the degeneracy of the unphysical sector are properly separated. 

Another issue awaiting for careful consideration is the return to real time and Minkowski 
space-time. The renormalization group studies of the Euclidean theory can shed
light on the way the contributions of the off-shell modes pile up as the long distance
physics is approached. But confinement of color is beyond this issue, it concerns
the dynamics of modes on the mas-shell. A basic element of the renormalization group
idea is the successive dealing with the degrees of freedom. The order of their
elimination is in principle arbitrary but it is advised to start with simple,
perturbative modes and finish with soft, large amplitude fluctuations. In fact,
the piling up the informations gained during the elimination process makes the 
effective dynamics better prepared to deal with the non-perturbative modes at a later
stage of the elimination process. Recall the renormalization group
approach to fermions at finite density where the blocking zooms into the Fermi sphere
in the Brioullin zone instead of the zero momentum point as for bosonic particles. In order to
address the confinement problem in Yang-Mills theory with the renormalization group method 
we have to zoom into the mass-shell which is possible in Minkowski space-time only. 
There is reason to suspect that mass-shell singularities are stronger in the Yang-Mills 
vacuum than for non-confining models. For instance the perturbative collinear divergences are
stronger for non-Abelian gauge theories and the view of confinement
as an Anderson localization in space-time\cite{laurent} suggests that pinch-singularities may arise, too.
The liquid models of the vacuum indicate the presence of an unusually large number
of soft modes which enhance the dressing, as well.

Finally, the truncation of the ansatz used in solving the evolution equation may
be critical for models with condensate. The (approximate) dynamical Maxwell-cut 
can not be obtained in the scalar $\phi^4$ theory or the sine-Gordon model when 
the loop corrections are collected in a truncated power or Fourier series 
representation of the local potential. In a similar manner one expects the need
of more flexible ansatz than those what has been used so far in Yang-Mills models
to address the issue of degeneracy.

\section{Summary}
Seemingly disparate points are related in this work. The precursor of the
formation of a condensate, a large degree of degeneracy when the vacuum is
constructed by the successive turning on the modes in the plane wave basis
is claimed to be related to gluon confinement. Even if correct, this view
does not add much to our understanding of color confinement, it rather orients
our attention to some difficulties waiting us along the road. Gauge fixing is an
ever returning problem for non-perturbative methods for Yang-Mills models. The
Wick rotation and the choice of the ansatz are important problems of the
functional renormalization group method independently of their possible
role in Yang-Mills theories. Their improvement would be a gain for other
domains, too.

Finally, another subjective remark about the functional renormalization group.
I think that it is a promising method whose limitation is not yet in sight. It can conveniently interpolate between
numerical and more intuitive, analytically based schemes to solve strongly
coupled theories. As such, it should ultimately integrate into itself the experiences
gained in lattice gauge theory and use them where this latter is not reliable,
in real time scattering processes and finite particle density.

\section*{Acknowledgments}
I thank Jean Alexandre and Vincent Pangon for the opportunity to work with them
on the nice problems\cite{jean,grg,vincentphi,vincentsg} covered in this paper.

\end{document}